\documentstyle[prl,aps,epsf,eqsecnum,multicol]{revtex} 
\begin{document} 
\bibliographystyle{prsty}
\title{Hydrodynamics of liquids of arbitrarily curved flux-lines and 
vortex loops} 
 
\author{Panayotis Benetatos\footnote{Present address: Physics Department, Harvard University, Cambridge, MA 02138} and M. Cristina Marchetti} 
\address{Physics Department, Syracuse University, Syracuse, NY 13244} 
 
\date{\today} 
 
\maketitle

\begin{abstract} 
We derive a hydrodynamic
model for a liquid of arbitrarily curved flux-lines and vortex loops using
the mapping of the vortex liquid onto a liquid of {\it relativistic} charged
quantum bosons in 2+1 dimensions recently suggested by 
Te{\v s}anovi{\' c} and by Sudb{\o} and collaborators. 
The loops in the flux-line
system correspond to particle-antiparticle fluctuations in the bosons. 
We explicitly
incorporate the externally applied magnetic field which in the boson
model corresponds to a chemical potential associated with the conserved
{\it charge} density of the bosons. We propose this model as a convenient
and physically appealing starting point for studying the properties of the
vortex liquid.
\end{abstract} 
 
\begin{multicols}{2}

\section{Introduction}

The physics of vortex matter has been a very active field of research since the
discovery of high-temperature superconductivity \cite{Blatter,CrabtreeNelson}. The melting of the Abrikosov vortex lattice at a field well
below the mean field upper critical line, $H_{c2}(T)$, is now well established
both theoretically and experimentally. The nature of the resulting vortex
liquid phase remains, however, unclear. Various approaches have been used to
study the properties of the liquid, ranging from the mapping of the statistical
mechanics of vortex lines onto that of (nonrelativistic) two-dimensional
quantum bosons \cite{Nelson88,MPA89,seung,NelsonVin}, to continuum hydrodynamic models
\cite{mcmnelphysica,paper1,mcmdrn00}. Recently, it has been
proposed that the vortex liquid phase should be viewed not as a collection of
well-defined directed vortex lines induced by the externally applied field, but
rather as a system where closed vortex loops proliferate, with components in
all directions \cite{Tesanovic1,Tesanovic2}. This picture has been substantiated by numerical
work by Sudb{\o} and collaborators \cite{sudbo0,sudbo1,sudbo2,sudbo3,sudbo4,sudbo5,sudbo99}. Such a vortex liquid has
been described by Kiometzis et al. \cite{Kiometzis} and by Te{\v s}anovi{\' c}
\cite{Tesanovic1,Tesanovic2} by mapping it onto a system of {\it relativistic}
two-dimensional quantum bosons, where the proliferation of vortex loops and
overhangs in the flux-line system corresponds to particle-antiparticle
creation and annihilation events in the bosons. This mapping has been developed in some detail for
the case of zero external field.

In this paper we consider a vortex-line liquid in an arbitrary homogeneous
external field, ${\bf H}$. The only restriction is that $H$ be well below
the mean field upper critical line, $H_{c2}$, 
so that the London approximation applies. The liquid in general consists of both field-induced
vortices (on the average aligned with the external field, but where large
fluctuations leading to overhangs are not excluded) and closed vortex loops
generated spontaneously by thermal fluctuations. On the basis of general considerations, the
statistical mechanics of such a vortex liquid maps onto that of {\it
relativistic} charged quantum bosons in $2+1$ dimensions coupled to a screened
electromagnetic field. Our model generalizes that of Te{\v s}anovi{\' c} \cite{Tesanovic2} in
that it explicitly incorporates the external field ${\bf H}$ which in the boson
model corresponds to a chemical potential, $\mu_r$, associated with the conserved
{\it charge} density of the bosons. By manipulating the relativistic boson
action, we then derive a hydrodynamic model for a liquid of arbitrarily curved
flux-lines and closed vortex loops. We propose this model as a convenient and physically appealing starting point for studying the properties of the vortex liquid.

We consider for simplicity a clean, isotropic type-II superconductor in the
mixed state. The anisotropic case is discussed in the Appendix.
In the London limit, the familiar Ginzburg-Landau free energy
functional can be rewritten in terms of discrete vortex degrees of freedom by
parametrizing  the $i$-th line by its position ${\bf r}_i(s_i)$, with $s_i$ the
arclength along its curve, as \cite{barford88,MCM91,BrandtRev}

\begin{eqnarray}
\label{fendis}
{\cal G}\big[\{{\bf r}_i(s_i)\}, {\bf
H}\big]&\;=\;&U_{self}\;+\;U_{int}\nonumber\\
& &\;-\;\frac{\phi_0}{4\pi}\sum_{i=1}^N\int d{\bf r}_i\cdot{\bf H}\;, 
\end{eqnarray}
where
\begin{equation}
\label{term1}
U_{self}\approx\epsilon_1 \sum_{i=1}^N \int ds_i 
\end{equation}
is an approximate form for the self-energy of a single vortex line, and
\begin{equation}
\label{term2}
U_{int}={1\over 2}\sum_{i\not = j}\int d{\bf r}_i \cdot\int' d{\bf r}_j V(|{\bf r}_i-{\bf r}_j|)\;,
\end{equation}
with $V(r)=(\epsilon_0/r) e^{-r/\tilde\lambda}$, describes the screened
interaction between vortex segments on different lines.  Here
$\epsilon_0={\phi_0^2}/({4\pi \tilde \lambda})^2$ and
$\epsilon_1=\epsilon_0\ln\kappa$. Also, $\;\tilde \lambda$ is the effective London
penetration depth and $\kappa \equiv \tilde \lambda/\xi$, with $\xi$ the
coherence length. The prime on 
the integral sign indicates that the interaction must be cutoff at 
intervortex separations of the order of $\xi$. The last term in Eq. (\ref{fendis})
represents the interaction with the homogeneous external field, ${\bf H}$. A
short distance cutoff $\approx\xi$  has been assumed to regularize the
integrals at short scales. The canonical partition function of the system is 
\begin{equation}
\label{partf}
{\cal Z}({\bf H},T)=\int' {\cal D}\{{\bf r}_i(s_i)\}e^{-{\cal G}/k_BT}\;.
\end{equation}
The prime over the integral sign indicates that the functional integration must
be performed over all flux-line configurations $\{{\bf r}_i(s_i)\}$ subject to
the constraint ${\bf \nabla}\cdot {\bf B}=0\;$ (i.e. no flux-line can start or
stop within the sample). The expression for the vortex
energy given in Eqs. (\ref{fendis}{--}\ref{term2}) is the starting point of much theoretical work on vortex arrays. As required, the free energy
is a scalar and it is therefore rotationally invariant.

A  familiar approximation to the general free energy given in
Eq. (\ref{fendis}) can be obtained at low fields and temperatures, where
field-induced vortex lines interact weakly and fluctuations away from the external field
direction are small. In this case it is convenient to choose the $z$ axis in
the direction of the applied field, ${\bf H}={\bf \hat{z}}H_0$ and to
parametrize the vortex positions as ${\bf r}_i=\big[{\bf r}_{\perp
i}(z),z\big]$. Asuming small transverse fluctuations of the vortices and
retaining only the interaction among vortex segments at the same ``height'' $z$ (to be referred to as the {\it local} approximation), one obtains
\begin{eqnarray}
\label{smalltilt}
{\cal G}\big[\{{\bf r}_i(s_i)\}, H_0& & \big]\;\approx
\sum_{i=1}^N\int dz\bigg\{{\epsilon_1\over 2}\Big[{d{\bf r}_{\perp i}(z)\over dz}\Big]^2-\mu\bigg\}\nonumber\\
& & +{1\over 2}\sum_{i\not = j}\int dz V(|{\bf r}_{\perp i}(z)-{\bf r}_{\perp j}(z)|)\;.
\end{eqnarray}
with $\mu={H_0\phi_0}/{(4\pi)}-\epsilon_1\;$ \cite{Nelson88,seung}.
Eq. (\ref{smalltilt}) is clearly
only appropriate to describe field-induced {\it directed} lines and it neglects
the possibility of thermally-induced closed vortex loops. It is well known that
the partition function of the array of directed vortex lines described by
Eq. (\ref{smalltilt}) maps onto the quantum  mechanical partition function in
the path integral formulation for a gas of two-dimensional bosons
\cite{Nelson88,seung,mfisher,NelsonJStatPhys,TN}. In this
mapping, the vortex lines correspond to the boson world lines, $\epsilon_1$ is the boson mass, and $z$ represents the imaginary time. This boson mapping has been exploited by Nelson and collaborators to study the properties of vortex liquids and is particularly useful for the description of vortex arrays pinned by 
correlated disorder \cite{NelsonVin}. The approximate vortex free energy given in
Eq. (\ref{smalltilt}) is invariant under a uniform
tilt of the external field  away from the $z$ direction. In the boson analogy, 
this translates into invariance with respect to Galileian boosts \cite{TNnote}.

One of the limitations of this boson model is that it neglects the nonlocality of the intervortex 
interaction in the external field direction. Feigel'man and collaborators  
argued that this nonlocality can be incorporated in the boson formalism  
by mapping the vortex partition function onto the path integral  
of two-dimensional {\it charged} bosons \cite{charge_note}  coupled to a screened gauge field 
that mediates the non-instantaneous interaction among the bosons
\cite{feigelJETP,feig}. 
The boson action proposed by these authors suffers, however, from one crucial problem {---}  
it does not incorporate the symmetry of the original vortex 
free energy given in Eq. (\ref{fendis}). More precisely, the boson analogue 
of a rotation of the three-dimensional space where the vortices are embedded is a Lorentz  
transformation of the $2+1$-dimensional space-time where the bosons are embedded. 
Since the original vortex energy is rotationally invariant, its dual boson 
action should be Lorentz invariant. But the model implemented by 
Feigel'man and collaborators lacks a definite symmetry as 
the field and the interaction part of the Lagrangian density are Lorentz invariant, 
but the approximate form used for the 
free-particle part is only Galileian invariant. 
We note that the non-Gaussian hydrodynamic model proposed recently by us \cite{paper1} suffers 
from the same problem, as it was actually {\it derived} from the nonrelativistic charged boson model. 
This problem is in retrospect immediately apparent in the hydrodynamic formulation, 
where the free energy can be written in terms of the coarse-grained fields 
$\big[\hat{n}^H({\bf r}), \hat{\bf t}^H({\bf r})\big]$ as the sum of three terms, as in Eq. (\ref{fendis}), 
\begin{eqnarray} 
\label{fenhyd} 
{\cal G}^H\big[\hat{n}^H, \hat{\bf t}^H,
H_0\big]\;=\;U^H_{self}\;+\;U^H_{int}
\;-\;NLH_0\frac{\phi_0}{4\pi}\;, 
\end{eqnarray} 
with 
\begin{equation} 
\label{term1H} 
U^H_{self}=NL\epsilon_1 +\frac{1}{2}\int_{\bf r}\epsilon_1\frac{|\hat{\bf t}^H({\bf 
r})|^2}{\hat{n}^H({\bf r})} 
\end{equation} 
and 
\begin{eqnarray} 
\label{term2H} 
U^H_{int}=\frac{1}{2n_0^2} \int_{\bf r} \int_{{\bf r}'} & & \Big[ 
K_t ({\bf r} - {\bf r}') \hat{\bf t}^H({\bf r}) \cdot \hat{\bf t}^H({\bf r}')\nonumber\\ 
& &\;+\;K_c({\bf r} - {\bf r}')\hat{n}^H ({\bf 
r})\hat{n}^H ({\bf r}')\Big] \;. 
\end{eqnarray} 
Here $K_c({\bf r})$ is the real space compressional modulus 
and $K_{t}({\bf r})$  is the compressional part of the real space tilt 
modulus. For an isotropic superconductor, $K_c({\bf r})=K_{t}({\bf r})$.  
Since  $[\hat{n}^H({\bf r}),\hat{\bf t}^H({\bf r})]$ is  
a vector under rotations (proportial to the local induction ${\bf B}$ 
in the superconductor, with $\hat{n}^H({\bf r})\sim B_z({\bf r})$ 
and $\hat{\bf t}^H({\bf r})\sim {\bf B}_\perp({\bf r})$), we see that, for $L\rightarrow \infty$, 
$U_{int}^H$ is a scalar and therefore rotationally invariant, 
while $U_{self}^H$ is not a scalar and therefore is not rotationally
invariant. Note that, in contrast, both the uncharged boson model \`a la Nelson \cite{TN}
and conventional Gaussian hydrodynamics \cite{mcmnelhyd,mcmnelphysica}
(where the field $\hat{n}^H({\bf r})$ in the denominator 
of Eq. (\ref{term1H}) is replaced by its equilibrium value) are consistent approximations to 
the free energy of directed vortex lines and respect the restricted 
symmetry under global rotations about the $z$ axis. 
These observations indicate that, as recently pointed 
out in Refs. 
\cite{Kiometzis,Tesanovic1,Tesanovic2,sudbo0,sudbo1,sudbo2,sudbo3,sudbo99}, the proper two-dimensional boson model 
dual to the original three-dimensional vortex problem is a gas 
of {\it relativistic} charged bosons in $2+1$-dimensional space-time, 
interacting via a screened electromagnetic field.

In this paper, starting from the action of  relativistic charged quantum bosons, we derive a nonlocal, 
nonlinear hydrodynamic model of liquids of arbitrarly curved vortex lines 
that properly incorporates the full rotational symmetry of the vortex energy. 
As discussed in Refs. \cite{Kiometzis,Tesanovic1,Tesanovic2,sudbo0,sudbo1,sudbo2,sudbo3,sudbo4,sudbo5,sudbo99},  
this leads to some interesting new concepts, namely, the appearence of thermally excited {\it antivortex} 
lines and {\it vortex loops}. It is well known that a loop gas is dual to a gas 
of relativistic bosons in $2+1$ space-time dimensions.  
This duality has been exploited recently by a few authors to study the properties of  
vortex liquids in type-II superconductors.  
Kleinert and collaborators 
\cite{Kiometzis,Kleinertbook} 
have developed a field theory of a gas of fluctuating closed  
vortex loops and have applied renormalization group ideas to study its phases. 
Te{\v s}anovi{\' c} has proposed a theory \cite{Tesanovic2} which separates 
``field-induced'' from ``thermally-generated'' degrees of freedom. The 
former correspond to directed flux-lines, while the latter correspond to 
closed vortex 
loops. A prediction of this work is the existence of a phase transition, 
dubbed the 
$\Phi$-transition, within the flux liquid phase from a low temperature liquid 
of lines with finite tension to a liquid phase that 
lacks superconducting coherence in all directions and where 
vortex loops proliferate yielding the vanishing of the flux-line tension.  
Numerical evidence for this $\Phi$-transition has recently come from the numerical 
simulations of Sudb{\o} and 
collaborators \cite{sudbo99}. 

In this paper, we propose a long 
wavelength description of a liquid of interacting field-induced vortex lines and vortex loops 
which treats both {\it 
on equal footing}. The external magnetic field is incorporated explicitly and 
controls the density of field-induced lines. 
In the dual description it enters as the chemical potential of the bosons, in analogy to the  
nonrelativistic boson model. The important difference here is that the chemical potential 
couples to the local {\it charge} density of bosons, which is the relevant
conserved variable as contrasted to the particle number density. 
The hydrodynamic model proposed here  is physically appealing model and 
naturally generalizes the familiar Gaussian hydrodynamics. 
Its ramifications and particularly its standing with respect 
to the aforementioned $\Phi$-transition are interesting directions for 
future work.

\section{Derivation of the hydrodynamics of flux-lines and flux-loops from 
a relativistic boson analogy} 
 
In this section, we use the methods discussed  in Ref. \cite{paper1} to derive the
hydrodynamic free energy  of arbitrarly curved vortex lines. As usual, the
vortex lines are viewed as imaginary-time  world lines of $2D$ bosons. The need
for a relativistic description is immediately apparent by considering the
self-energy part of the vortex energy, given in Eq. (\ref{term1}), where the
arclength of a curve segment can be parametrized as $ds=\sqrt{(dz)^2+(d{\bf
r}_\perp)^2}$. Eq. (\ref{term1}) is formally identical to the imaginary-time
action of a relativistic free particle of rest mass
$m\leftrightarrow\epsilon_1$ \cite{CTF},
\begin{equation}
\label{relact}
S=-mc\int_{s_1}^{s_2}ds
 =-mc\int_{s_1}^{s_2}\sqrt{c^2(dt)^2-(d{\bf r}_{\perp})^2}\;,
\end{equation}
where $s_1$ and $s_2$ are two space-time events and 
$c\tau \leftrightarrow z$, with $\tau\equiv it$  the imaginary time. 
The speed of light $c$ does not have a counterpart in the flux-line free energy, 
but is explicitly kept in this section  
to enable us to obtain the nonrelativistic limit. 
The correspondence between boson and vortex variables  
is summarized in Table 1. One important diference between our mapping
and the mapping to $2D$ {\it nonrelativistic} bosons discussed extensively
in the literature is that here the boson mass, $m_b$, corresponds
to the vortex line energy rather than to the titl energy per unit length.
This distinction becomes important in the case of anisotropic
materials, as discussed in the Appendix.
\vspace{0.2in} 
\begin{table} 
\caption{\label{table1} Correspondence between boson and vortex variables.} 
\begin{tabular}{cccc}  
&bosons & vortices &\\ \tableline\\
&$c\tau$ & $z$&\\ \\ 
&$\beta\hbar c$ & $L$ &\\ \\ 
&$S/\hbar$ & ${\cal F}/k_BT$&\\ \\ 
&$m_bc^2/\hbar c$ & $\epsilon_1/k_BT$&\\ \\ 
&$e_b^2/\hbar c$ & $4\pi\epsilon_0/k_BT$&\\ \\ 
&$\mu/\hbar c$ & $H_0\phi_0/(4\pi k_BT)$&\\ \\ 
&$\lambda_b$ & $\tilde\lambda$& \\ \\ 
&${\bf t}_b/c$ & ${\bf t}\approx{\bf B}/\phi_0$ &\\\\ 
\end{tabular} 
\end{table} 
\vspace{0.2in} 

As shown below, the vortex lines can be interpreted as the world lines of {\it relativistic}  
spin-$0$, charged bosons  
in $2+1$ space-time dimensions, 
interacting via a screened electromagnetic interaction \cite{bardakci}.  
By applying standard methods \cite{Kapusta1,Kapusta2} 
of rewriting the boson Hamiltonian in the language of  
second quantization and transforming to coherent states,  
the imaginary-time path integral representation 
for the grand-canonical partition function of the bosons  
is written as 
\begin{equation} 
Z_G^{\rm bos}=\int_{\Phi^*({\bf r}_\perp,\beta\hbar)=\Phi^*({\bf r}_\perp,0) 
        \atop\Phi({\bf r}_\perp,\beta\hbar)=\Phi({\bf r}_\perp,0)} 
  {\cal D}\Phi{\cal D}\Phi^*{\cal D}{\bf a}{\cal D}{\bf A} 
  e^{-{\cal S}_r[\Phi, \Phi^*, {\bf a}, {\bf A}]/\hbar}, 
\end{equation} 
with the action 
\end{multicols} 
\begin{eqnarray} 
\label{Rel} 
{\cal S}_r[\Phi, \Phi^*, {\bf a}, {\bf A}]= & &\int_0^{\beta \hbar} d\tau  \int d^2{\bf 
r}_{\perp}\bigg\{\frac{m_bc^2}{2}\Phi\Phi^*  
+\frac{1}{2m_bc^2}\big[(-i\hbar\partial_\tau+e_ba_0+i\mu_r)\Phi\big] 
      \big[(i\hbar\partial_\tau+e_ba_0+i\mu_r)\Phi^*\big]\nonumber\\ 
& & +\frac{1}{2m_b}|(-i\hbar{\bf \nabla}_\perp+\frac{e_b}{c}{\bf a}_\perp)\Phi|^2 
+ {\cal L}_F[{\bf a}, {\bf A}]\bigg\}\;, 
\end{eqnarray} 
where  
\begin{eqnarray} 
\label{Lfield} 
{\cal L}_F[{\bf a}, {\bf A}]= \frac{1}{8\pi}\bigg\{({\bf 
\nabla}\times{\bf a})^2 
+\frac{2i}{\lambda_b}({\bf 
\nabla}\times{\bf a})\cdot{\bf A} 
+({\bf \nabla}\times{\bf A})^2\bigg\}\;  
\end{eqnarray} 
\begin{multicols}{2} 
\noindent is the free field Lagrangian 
density and $\nabla\equiv\big({1\over c}\partial_\tau,\nabla_\perp\big)$.  
Here, $\Phi({\bf r}_{\perp}, t)\;$  
is a complex scalar matter field that 
describes bosons with positive and negative charge, with  
$e_b$ the boson charge.  
The world lines of 
bosons of opposite charge, which are {\it antiparticles} to each other,  
correspond to flux-lines with 
opposite vorticity. The constraint on the fields $\Phi$ and $\Phi^*$ comes from the permutation 
symmetry requirement and reflects boson statistics. 
The boson field $\Phi$ is coupled to a {\it massive} electromagnetic  
field ${\bf a}=\big(a_0,{\bf a}_\perp\big)$ and the gauge field  
${\bf A}=\big(A_0,{\bf A}_\perp\big)$ provides the screening. 
Integrating out the field ${\bf A}$ in Eqs. (\ref{Rel}) and (\ref{Lfield}) under the gauge $\nabla \cdot {\bf a}=0$
gives a mass term $|{\bf a}|^2/(8\pi\lambda_b^2)$. 
Under Lorentz transformations, the field $\Phi$ is a scalar and the boson
action has full Lorentz invariance. The action also respects a global $U(1)$ 
symmetry, $\Phi\rightarrow\Phi'=\Phi e^{i\alpha}$, which leads to the conserved current, 
${\bf j}=(j_0,{\bf j}_\perp)$, given by  
%
\begin{eqnarray} 
\label{current} 
{\bf j}=\frac{e_b}{2m_b}\big[\Phi^*\big( -i\hbar\nabla 
    +\frac{e_b}{c}{\bf a}\big)\Phi 
         +\Phi\big( i\hbar\nabla+\frac{e_b}{c}{\bf a}\big)\Phi^*\big], 
\end{eqnarray} 
%
with $\nabla\cdot{\bf j}=0$. The temporal component 
$j_0/c$ of the 2+1-current is the charge density, which is the appropriate 
conserved quantity in a relativistic theory. 
Finally, we have introduced a  
chemical potential, $\mu_r$, which couples to the conserved {\it charge} 
density.  
This should be contrasted to the nonrelativistic boson model which maps onto a liquid of {\it directed} vortex lines and where 
the relevant conserved quantity is particle number. 
The quantum relativistic 
model naturally incorporates spontaneous creation and annihilation of 
particle-antiparticle pairs. The flux-line analogue of this is the  
creation of oriented  {\it vortex loops}.  
It will become apparent below that the chemical potential 
corresponds to the external field ${\bf H}$.

In order to highlight the connection with the nonrelativistic boson model  
used by Nelson and collaborators, we rewrite 
the boson field in terms of an amplitude and a phase, $\Phi({\bf 
r}_{\perp}, \tau)=\sqrt{\rho({\bf 
r}_{\perp}, \tau)}\exp{[i\theta({\bf 
r}_{\perp}, \tau)]}\;$, with the result 
\end{multicols} 
\begin{eqnarray} 
\label{Rel2} 
{\cal S}_r[\rho, \theta, {\bf a}, {\bf A}]=& &\int_0^{\beta \hbar}d \tau  \int d^2{\bf 
r}_{\perp}\bigg\{\frac{m_bc^2}{2}\rho + 
\frac{\hbar^2}{8m_b\rho}(\nabla\rho)^2 
+\frac{\rho}{2m_b}(\hbar{\bf 
\nabla}_{\perp}\theta+\frac{e_b}{c}{\bf
a}_{\perp})^2+\frac{\rho}{2m_bc^2}(\hbar\partial_{\tau}\theta+e_ba_0+i\mu_r)^2
\nonumber\\
& &+{\cal L}_F[{\bf a}, {\bf A}]\bigg\}\;. 
\end{eqnarray} 
\begin{multicols}{2} 
 
A boson condensate phase is signalled by a macroscopic occupation $\rho_c$ 
of the ${\bf p}=0$ momentum 
state, defined in terms of the average order parameter as 
\begin{equation} 
\rho_c=|\langle\Phi({\bf r}_\perp,\tau)\rangle|^2\;. 
\end{equation} 
Notice that in general $\rho_c\not=\rho_0\equiv\langle\rho({\bf r}_\perp,\tau)\rangle= 
\langle|\Phi({\bf r}_\perp,\tau)|^2\rangle$, although the two quantities are 
equal within the mean field  approximation described below. 
 
In the spirit of Landau-Ginzburg mean field theory, we evaluate the partition function by the method  
of steepest descent. The stationarity condition gives nonlinear 
differential equations for the various fields. Restricting ourselves to  
solutions which are stationary and spatially homogeneous, 
the extrema conditions yield the equations, 
\begin{eqnarray} 
& & \frac{\delta S_r}{\delta\rho}=\frac{m_bc^2}{2}+\frac{e_b^2|{\bf a}_\perp|^2}{2m_bc^2} 
       +\frac{1}{2mc^2}(e_ba_0+i\mu_r)^2=0\;,\\ 
& & \frac{\delta S_r}{\delta{\bf a}_\perp}=\frac{e_b^2}{m_bc^2}\rho{\bf a}_\perp 
    +\frac{{\bf a}_\perp}{4\pi\lambda_b^2}=0\;,\\ 
& & \frac{\delta S_r}{\delta a_0}=\frac{e_b}{m_b c^2}\rho(e_b a_0+i\mu_r) 
    +\frac{a_0}{4\pi\lambda_b^2}=0. 
\end{eqnarray} 
One finds of course ${\bf a}_\perp=0$, corresponding to the absence 
of charge current in the homogeneous state.  
A spatially homogeneous saddle point solution with $|\langle\Phi\rangle|^2=\rho_0\not=0$  
exists only for $|\mu_r|>m_bc^2$ \cite{foot_mu}. 
The solution is given by 
\begin{eqnarray} 
\label{rho0} 
& & \rho_0=\frac{|\mu_r|-m_bc^2}{4\pi\lambda_b^2e_b^2}\;,\;\;\;\;{\rm for}\;\;|\mu_r|> m_bc^2\;,\nonumber\\ 
& & \rho_0=0\;,\;\;\;\;{\rm for}\;\;|\mu_r|\leq m_bc^2\;. 
\end{eqnarray} 
Within this mean field approximation, the occupation of the zero momentum 
state is $\rho_c = \rho_0$. 
The free energy in the condensate phase of bosons 
is given by 
\begin{equation} 
F_G^c(T,\Omega,\mu_r)=-k_BT\ln{\cal Z}_G^c=-\hbar \Omega\frac{(m_bc^2-\mu_r)^2} 
   {8\pi\lambda_b^2e_b^2}\;, 
\end{equation} 
where $\Omega$ is the area of the $2D$ boson gas. 
The mean charge density, $\langle j_0/c\rangle$,  in this mean-field approximation 
is  
\begin{equation}
\langle j_0/c\rangle=-\frac{1}{\Omega} \Big(\frac{\partial F_G^c}{\partial\mu_r}\Big)_{T,\Omega}=\pm ie_b\rho_0\;,
\end{equation}
where the positive (negative) sign should be chosen 
for $\mu_r>0$ ($\mu_r<0$). 
For charged bosons the chemical potential is associated with the electric charge 
and allows for a positive (or negative)  
average charge density. Antiparticles  
have a charge and a chemical potential opposite to that of particles. 
 
Within mean field theory, all the charge is in the condensate phase. 
When fluctuations are incorporated in the theory, the mean charge density, $\langle\rho\rangle$, 
differs from the condensate fraction, $\rho_c$, 
even at zero temperature. 
 
 The nonrelativistic limit for the action can be obtained by 
shifting  the the chemical potential by 
$m_bc^2\;$,
\begin{equation} 
\label{mushift} 
\mu_r=\mu+m_bc^2\;,
\end{equation} 
where $\mu$ is the nonrelativistic chemical potential \cite{foot_mu}. 
The action becomes 
\end{multicols} 
\begin{eqnarray} 
\label{nrellim} 
{\cal S}_r[\rho, \theta, {\bf a}, {\bf A}]&=&\int_0^{\beta \hbar}d\tau \int d^2{\bf 
r}_{\perp}\bigg\{ 
\frac{\hbar^2}{8m_bc^2\rho}(\partial_{\tau}\rho)^2+\frac{\hbar^2}{8m_b\rho}({\bf 
\nabla}_{\perp}\rho)^2 +i\rho(\hbar\partial_{\tau}\theta+e_ba_0+i\mu)\nonumber\\ 
& &+\frac{\rho}{2m_b}(\hbar{\bf 
\nabla}_{\perp}\theta+\frac{e_b}{c}{\bf a}_{\perp})^2+\frac{\rho}{2m_bc^2}(\hbar\partial_{\tau}\theta + e_ba_0+i\mu)^2 
+{\cal L}_F[{\bf a}, {\bf A}]\bigg\}\;. 
\end{eqnarray} 
\begin{multicols}{2} 
\noindent The nonrelativistic boson model of T{\"a}uber and Nelson \cite{TN} is then recoverd by letting 
$c\rightarrow\infty$. In this limit the 2D bosons only interact with the scalar  
field $a_0$, and there is no magnetic interaction, 
\end{multicols} 
\begin{eqnarray} 
\label{nrellim_answer} 
{\cal S}_{nr}[\rho, \theta]&=& 
   \int_0^{\beta \hbar}d\tau \int d^2{\bf r}_{\perp} 
\bigg\{ 
   i\hbar\rho\partial_{\tau}\theta+i\rho e_ba_0-\rho\mu
   +\frac{\hbar^2}{8m_b\rho}({\bf\nabla}_{\perp}\rho)^2 
   +\frac{\rho}{2m_b}(\hbar{\bf\nabla}_{\perp}\theta)^2 \nonumber\\
& &  +\frac{1}{8\pi}\big[({\bf \hat{z}}\times\nabla_\perp) a_0\big]^2 
   +\frac{a_0^2}{8\pi\lambda_b^2} 
\bigg\}\;. 
\end{eqnarray} 
\begin{multicols}{2} 
\noindent The scalar potential $a_0$ mediates the instantaneous screened Coulomb 
interaction. By integrating it out  
we recover the familiar {\it instantaneous} interaction term with Fourier components 
$|\rho({\bf q})|^2 4\pi\lambda_b^2 e_b^2/2(1+q_{\perp}^2\lambda_b^2)$. 
 
In order to obtain the hydrodynamic free energy, we proceed as in Ref. \cite{paper1} 
and perform a Hubbard-Stratonovich transformation  
to eliminate the boson amplitude and phase fields in favor of familiar hydrodynamic fields. 
First we eliminate the phase $\theta$ in Eq. (\ref{Rel2}) in favor of 
a new vector field ${\bf P}$, with the result 
\end{multicols} 
\begin{eqnarray} 
\label{PRel} 
{\cal S}'_r[\rho, {\bf P}, {\bf a},& & {\bf A}]= 
   \int_0^{\beta \hbar}d\tau \int d^2{\bf r}_{\perp}\bigg\{ 
       \frac{m_bc^2}{2}\rho 
       +\frac{\hbar^2}{8m_bc^2\rho}(\partial_{\tau}\rho)^2 
       +\frac{\hbar^2}{8m_b\rho}({\bf \nabla}_{\perp}\rho)^2  
       +\frac{\rho}{2m_b}\Big[\frac{P_z^2}{c^2}+{\bf P}_{\perp}^2\Big]\nonumber\\ 
 & &   +i\frac{\rho P_z}{m_bc^2}\big(e_ba_0+i\mu_r\big) 
       +\frac{ie_b\rho}{m_bc}{\bf P}_{\perp}\cdot{\bf a}_{\perp} 
       +{\cal L}_F[{\bf a}, {\bf A}] 
       +\frac{\rho_0 \hbar^2}{m_b}\ln{(\frac{\rho}{\rho_0})} 
   \bigg\} 
\end{eqnarray} 
\begin{multicols}{2} 
\noindent with the constraint 
\begin{equation} 
\label{consRel1} 
\frac{1}{c^2}\partial_{\tau}(\rho P_z)+{\bf \nabla}_{\perp}\cdot(\rho {\bf P}_{\perp})=0\;. 
\end{equation} 
Notice that the field ${\bf P}$ has {\it three} components. 
The temporal component $P_z$ is related to the conserved charge 
density of the theory, which in turn is related to the temporal 
variations of the phase.   
The last term in Eq. (\ref{PRel}) arises from the Jacobian of the functional integration and 
represents the nonlinear ``ideal gas'' part of the free energy. 
Finally, to make contact with the hydrodynamic fields used 
in our earlier work, we let 
\begin{equation} 
t_{bz}=\frac{\rho P_z}{m_bc}\;, 
\end{equation} 
\begin{equation} 
{\bf t}_{b\perp}=\frac{\rho {\bf P}_{\perp}}{m_b}\;, 
\end{equation} 
and integrate out the fields ${\bf a}$ and ${\bf A}$ to obtain 
an effective action 
\end{multicols} 
\begin{eqnarray} 
\label{tRel} 
{\cal S}'_r[\rho, {\bf t}_b]= 
  & &  \int_0^{\beta \hbar}d \tau  \int d^2{\bf r}_{\perp}\bigg\{ 
        \frac{m_bc^2}{2}\rho 
       +\frac{\hbar^2}{8m_bc^2\rho}(\partial_{\tau}\rho)^2 
       +\frac{\hbar^2}{8m_b\rho}({\bf \nabla}_{\perp}\rho)^2  
  +\frac{m_b}{2\rho }{\bf t}_b^2 
       -\frac{1}{c}\mu_r t_{bz}   
  \bigg\}\nonumber\\ 
   & &   + \frac{1}{2\Omega\beta\hbar}\sum_{\bf q}\frac{4\pi\lambda_b^2 e_b^2/c^2}{1+q^2\lambda_b^2} 
          |{\bf t}({\bf q})|^2\; 
\end{eqnarray} 
\begin{multicols}{2} 
\noindent with the constraint 
\begin{equation} 
\label{constraint} 
{\bf \nabla}\cdot {\bf t}_b=0\;, 
\end{equation} 
where ${\bf q}=(q_z/c, {\bf q}_{\perp})\;$. 
 
 
Since we are interested in the fluctuating vortices rather than in the 
relativistic bosons, we shall rewrite Eq. (\ref{tRel}) in the vortex 
language using the ``translation'' Table 1.  
We are interested in the 
thermodynamic limit where periodic or free boundary conditions give the 
same result and $\int_0^{\beta \hbar}cd \tau  \int d^2{\bf 
r}_{\perp}\rightarrow \int_{\bf r}=\int_0^L\int d{\bf r}_\perp$, 
where $L$ is the thickness of the sample.   
The resulting vortex free energy 
is given by 
\end{multicols} 
\begin{eqnarray} 
\label{vtRel} 
{\cal F}_r[{\bf t}, \rho]= 
  \int_{\bf r}\bigg\{ 
      \frac{\epsilon_1}{2}\rho  
     +\frac{(k_BT)^2}{8\epsilon_1\rho}({\bf \nabla}\rho)^2 
     +\frac{\epsilon_1}{2 \rho}{\bf t}^2 
      - \frac{H_0\phi_0}{4\pi}t_z \bigg\} 
   + \frac{1}{2\Omega L}\sum_{\bf q}\frac{4\pi\epsilon_0\tilde{\lambda}^2} 
        {1+q^2\tilde{\lambda}^2}|{\bf t}({\bf q})|^2\; 
\end{eqnarray} 
\begin{multicols}{2} 
\noindent with the constraint given by Eq. (\ref{constraint}) above. 
The fluctuating field ${\bf t}$ corresponds to 
the coarse-grained vortex density which determines the magnetic induction 
${\bf B}$ in the material. The chemical potential, $\mu_r$, of the 
relativistic bosons  
corresponds to the external applied field which gives rise to  
a net density of   
field-induced lines. 
The physical 
interpretation becomes clear by discussing the mean-field saddle 
point solution, 
obtained by applying the  
stationarity condition to the free energy, with the result 
\begin{eqnarray} 
\label{station1} 
& &\Bigg(\frac{\delta {\cal F}_r}{\delta {\bf t}_{\perp}}\Bigg)_{{\bf 
t}={\bf t}_0, \rho=\rho_0}=0\Longrightarrow {\bf t}_{\perp 0}=0\;, \\
\label{station2} 
& &\Bigg(\frac{\delta {\cal F}_r}{\delta \rho}\Bigg)_{{\bf 
t}={\bf t}_0, \rho=\rho_0}=0\Longrightarrow 
t_{z0}=\pm c\rho_0\;, \\
\label{station3} 
& & \Bigg(\frac{\delta {\cal F}_r}{\delta t_z}\Bigg)_{{\bf 
t}={\bf t}_0, \rho=\rho_0}=0 \Longrightarrow H_0= 
\pm\Big(H_{c1}+\phi_0\rho_0\Big)\;. 
\end{eqnarray} 
Conversely, the equilibrium solution is given by
\begin{eqnarray} 
& &\rho_0=\frac{|H_0|-H_{c1}}{\phi_0}\;,\;\;\; {\rm for}\;\; |H_0|>H_{c1},\nonumber\\ 
& &\rho_0=0\;,\;\;\;\;\;\;\;{\rm for} \;\;\;\;\;\;\;|H_0|\leq H_{c1}, 
\end{eqnarray} 
where $H_{c1}\epsilon_1 4\pi/\phi_0\leftrightarrow mc^2$. 
In other words, $\rho_0$ is a measure of the number density of {\it directed}  
field-induced vortices.  The field $\rho$ is a scalar and it is defined to be always positive 
(for simplicity, in the following we will consider the case of $H_0>0$, 
corresponding to the $+$ signs in Eqs. (\ref{station2}) 
and (\ref{station3}). 
The field $t_{z}$ is proportional to the $z$ component of the magnetic 
induction. Its equilibrium value in a spatially homogeneous system 
is simply proportional to $\rho_0$ because on large scales 
the contribution from vortices precisely cancels that of antivortices,  
and $t_{z0}=c\rho_0=cB_{z0}/\phi_0\;$.  
Locally, the fields $t_z$ and $\rho$ can, however, fluctuate independently, 
allowing for  
spontaneous vortex loop fluctuations, independent of the externally applied
field. As will be seen more explicitely in the next section, the field $\rho$ mediates 
the renormalization of the single-vortex stiffness due to such spontaneous loop 
fluctuations.  
 
The hydrodynamic free energy given in Eq. (\ref{vtRel}) provides 
a starting point for describing the  long wavelength properties 
of a liquid of interacting ({\it directed}) field-induced 
vortex lines and oriented vortex loops. It should be stressed 
that the distinction between directed lines and loops, 
while physically appealing, is strictly a single-line notion 
and loses much of its meaning in a continuum theory. 
At the level of hydrodynamics, spontaneous vortex loop fluctuations 
are incorporated via the field $t_z^L=t_z-\rho$.
As we will see below, one can construct an effective theory where  
loops are integrated out. Their role then enters as a renormalization of the vortex-line tension.

\section{Correlations} 
It is useful to evaluate the two-point correlation function 
of the hydrodynamic fields appearing in the free energy of Eq. 
(\ref{vtRel}). This is easily done within a Gaussian approximation 
for the free energy, obtained by introducing fluctuations  
of the fields  
about their equilibrium values, 
$\delta{\bf t}={\bf t}-{\bf t}_0$ and 
$\delta\rho=\rho-\rho_0$, and expanding the free  
energy to  
quadratic order in these fluctuations. 
For $\rho_0>0$, the Gaussian free energy is given by 
\end{multicols} 
\begin{eqnarray} 
\label{FH_Gaussian} 
{\cal F}_r^G[\delta{\bf t}, \delta\rho]=F_0+ 
     \frac{1}{2\Omega L}\sum_{\bf q}\bigg\{ & & 
     \Big[\frac{\epsilon_1}{\rho_0}+\frac{(k_BT)^2}{4\epsilon_1\rho_0}q^2\Big] 
                         |\delta\rho({\bf q})|^2 
     +\Big[\frac{\epsilon_1}{\rho_0}+V(q)\Big]|\delta{\bf t}({\bf q})|^2 
        \nonumber\\   
  & &   -\frac{\epsilon_1}{\rho_0}\big[\delta t_z({\bf q})\delta\rho(-{\bf q}) 
                                +\delta t_z(-{\bf q})\delta\rho({\bf q})\big] 
        \bigg\}\;, 
\end{eqnarray} 
\begin{multicols}{2} 
\noindent  
where averages over the free energy ${\cal F}_r^G$ are to be evaluated 
with the constraint 
\begin{equation} 
\label{constraint_q} 
i{\bf q}\cdot\delta{\bf t}({\bf q})=0\;. 
\end{equation} 
In Eq. (\ref{FH_Gaussian}), $F_0$ is the equilibrium value, 
$V(q)=4\pi\epsilon_0\tilde\lambda^2/(1+q^2\tilde{\lambda}^2)$, 
and $q^2=q_\perp^2+q_z^2$. 
It is instructive to integrate out $\delta\rho$ to obtain the effective free  
energy, 
\end{multicols} 
\begin{equation} 
{\cal F}^{\rm eff}_r[\delta t_z,{\bf t}_\perp]= 
      \frac{1}{2\Omega L}\sum_{\bf q}\bigg\{\bigg[ 
     \frac{(k_BT)^2q^2/4\rho_0}{\epsilon_1+(k_BT)^2q^2/4\epsilon_1} 
     +V(q)\bigg]|\delta t_z({\bf q})|^2 
     +\Big[\frac{\epsilon_1}{\rho_0}+V(q)\Big]|{\bf t}_\perp({\bf q})|^2     
  \bigg\} \;,
\end{equation} 
\begin{multicols}{2} 
\noindent which is written entirely in terms of fluctuations 
in the local induction, as at long wavelengths,
\begin{equation}
\delta{\bf B}\approx\phi_0\Big({\bf t}_\perp,\delta t_z\Big)\;.
\end{equation}
Finally, by separating ${\bf t}_\perp$ in its components longitudinal and 
transverse to ${\bf \hat{q}}_\perp={\bf q}_\perp/q_\perp$ according to 
\begin{equation} 
{\bf t}_\perp({\bf q})= 
   {\bf \hat{q}}_{\perp}t_\perp^{\rm L}({\bf q}) 
   +(\hat{z}\times\hat{\bf q}_\perp)t_\perp^{\rm T}({\bf q})\;, 
\end{equation} 
and using the constraint (\ref{constraint_q}) to eliminate 
$t_\perp^{\rm L}$ in favor of $\delta t_z$ we obtain, 
\end{multicols} 
\begin{equation} 
\label{FG_eff} 
{\cal F}^{\rm eff}_r[\delta t_z,{\bf t}_\perp]= 
      \frac{1}{2\Omega L}\sum_{\bf q}\bigg\{\bigg[ 
     \frac{\epsilon_1q_z^2}{\rho_0 q_\perp^2} 
       +\frac{\epsilon_1}{\rho_0}\frac{(k_BT)^2q^2/4\epsilon_1^2}{1+(k_BT)^2q^2/4\epsilon_1^2} 
     +V(q)\frac{q_z^2}{q_\perp^2}\bigg]|\delta t_z({\bf q})|^2 
     +\Big[\frac{\epsilon_1}{\rho_0}+V(q)\Big]|t^{\rm T}_\perp({\bf q})|^2     
  \bigg\}\;. 
\end{equation} 
\begin{multicols}{2} 
\noindent  
Comparing the effective free energy of Eq. (\ref{FG_eff})  
to the conventional Gaussian hydrodynamic free energy \cite{mcmnelhyd,mcmnelphysica}
obtained neglecting relativistic effects and given by 
\end{multicols} 
\begin{equation} 
{\cal F}^G[\delta t_z,{\bf t}_\perp]= 
      \frac{1}{2\Omega L}\sum_{\bf q}\bigg\{\bigg[ 
     \frac{\epsilon_1q_z^2}{\rho_0 q_\perp^2}+V(q)\frac{q_z^2}{q_\perp^2}\bigg]|\delta t_z({\bf q})|^2 
   +\Big[\frac{\epsilon_1}{\rho_0}+V(q)\Big]|t^{\rm T}_\perp({\bf q})|^2\bigg\} \;, 
\end{equation} 
\begin{multicols}{2}    
\noindent it is evident that ``relativistic'' effects 
yield short wavelength corrections to the single-vortex effective tension. 
Similarly, in the boson system they are responsible for corrections 
to the quasi-particle spectrum due to spontaneous  particle-antiparticle  
pair creation, Again, these effects are important only at finite  
wavevector.  
 
It is now straightforward to evaluate  
the Gaussian two-point correlation functions, 
with the result 
\end{multicols} 
\begin{eqnarray} 
\label{tztz} 
\langle \delta t_z({\bf q}) \delta t_z(-{\bf q})\rangle_G= 
   \frac{\rho_0k_BTq_\perp^2}{\rho_0V(q)q^2+\epsilon_1q_z^2 
  +\epsilon_1q_\perp^2\frac{(k_BT)^2q^2/4\epsilon_1^2}{1+(k_BT)^2q^2/4\epsilon_1^2}}\;, 
\end{eqnarray} 
\begin{eqnarray} 
\label{deltarhodeltarho} 
\langle \delta\rho({\bf q}) \delta\rho(-{\bf q})\rangle_G= 
     \frac{\rho_0k_BTq^2}{\rho_0V(q)q^2+\epsilon_1q_z^2 
  +\epsilon_1q_\perp^2\frac{(k_BT)^2q^2/4\epsilon_1^2}{1+(k_BT)^2q^2/4\epsilon_1^2}}\; 
            \frac{1+\rho_0V(q)/\epsilon_1}{1+(k_BT)^2q^2/4\epsilon_1^2}\;, 
\end{eqnarray} 
\begin{eqnarray} 
\label{deltarhotz} 
\langle \delta t_z({\bf q}) \delta\rho(-{\bf q})\rangle_G= 
     \frac{\rho_0k_BTq_\perp^2}{\rho_0V(q)q^2+\epsilon_1q_z^2 
  +\epsilon_1q_\perp^2\frac{(k_BT)^2q^2/4\epsilon_1^2}{1+(k_BT)^2q^2/4\epsilon_1^2}}\; 
            \frac{1}{1+(k_BT)^2q^2/4\epsilon_1^2}\;. 
\end{eqnarray} 
\begin{multicols}{2} 
The correlations of the in-plane part ${\bf t}_\perp$ of the tilt field are given by 
\begin{equation} 
\label{ttRel} 
\langle t_i (-{\bf q}) t_j ({\bf q}) \rangle_{G} = T^0_T ({\bf q}) P_{ij}^T({\bf q}_{\perp}) + 
T^0_L({\bf q}) P_{ij}^L({\bf q}_{\perp}) \; , 
\end{equation} 
with 
\begin{equation} 
\label{transversetiltRel} 
T^0_T ({\bf q}) = \frac{\rho_0k_B T}{\epsilon_1+\rho_0V(q)}\;, 
\end{equation} 
and 
\begin{equation} 
\label{longittiltRel} 
T^0_L({\bf q}) = \frac{q_z^2}{q_{\perp}^2}\langle t_z(-{\bf q}) t_z({\bf q})\rangle_G\;. 
\end{equation} 
$P_{ij}^L({\bf q}_{\perp})=\hat{q}_{\perp i}\hat{q}_{\perp j}$ and $P_{ij}^T({\bf q}_{\perp})=\delta_{ij}-P_{ij}^L({\bf q}_{\perp})$ are the in-plane longitudinal and transverse projection operators respectively.
To Gaussian order, the transverse part of the tilt field 
autocorrelator is the same as that obtained from the familiar hydrodynamics 
of directed lines.

The structure function, given by the  
autocorrelator of fluctuations in the vortex line density, $\delta t_z$, 
differs from its ``non-relativistic'' counterpart only at finite wave-vectors. In 
the hydrodynamic limit it simply reduces to the familiar result 
obtained using the 
(``non-relativistic'') hydrodynamics of directed lines, 
\begin{eqnarray} 
\label{tztz_hyd} 
\langle \delta t_z({\bf q}) \delta t_z(-{\bf q})\rangle_G\approx 
\frac{\rho_0k_BTq_\perp^2}{\rho_0V(q)q^2+\epsilon_1q_z^2}\;. 
\end{eqnarray} 

When ``relativistic effects'' are included the two  
fields $\rho$ and $t_z$ are no longer independent. It is then instructive to also consider  
fluctuations in a new field defined as their difference as $t_z^L=t_z-\rho$. 
This field fluctuates about a zero equilibrium value and vanishes 
identically when relativistic effects are neglected. 
Fluctuations in $t_z^L$ may be interpreted as a measure of fluctuations due 
to the spontaneous excitation of vortex loops. 
Its correlation function is given by 
\end{multicols} 
\begin{eqnarray} 
\label{tzLtzL} 
\langle \delta t^L_z({\bf q}) \delta t^L_z(-{\bf q})\rangle_G= 
     \frac{\rho_0k_BT}{\rho_0V(q)q^2+\epsilon_1q_z^2 
  +\epsilon_1q_\perp^2\frac{(k_BT)^2q^2/4\epsilon_1^2}{1+(k_BT)^2q^2/4\epsilon_1^2}}\; 
            \frac{q_z^2+\rho_0V(q)q^2/\epsilon_1}{1+(k_BT)^2q^2/4\epsilon_1^2}\;. 
\end{eqnarray} 
\begin{multicols}{2} 
 
The long wavelength limit of this correlation function  
is given by 
\begin{eqnarray} 
\lim_{{q_z}\rightarrow 0}\lim_{q_\perp\rightarrow 0}\langle\delta t^L_z(-{\bf q})\delta t^L_z({\bf q})\rangle_G 
     =\frac{k_BT}{\epsilon_1}\;, 
\end{eqnarray} 
irrespective of the order of the limits ($q_z\rightarrow 0$ first or $q_\perp\rightarrow 0$ first). 
In other words we have identified a correlation function that at long wavelengths 
yields the inverse single-line tension, $\epsilon_1$.  Calculating perturbative
corrections to this Gaussian correlator will be a way to probe the
renormalization of the single-line tension.

\section{Discussion}
 
We have presented a long wavelength (hydrodynamic) description of
a liquid of arbitrarly curved vortex line and loops that describes
on the same footing both field-induced and spontaneously generated vortices.
The long-wavelength model was obtained by exploiting the mapping of such 
a vortex liquid onto a gas of relativistic charged bosons in 2$D$ that was
recently discussed by Te{\v s}anovi{\' c} 
\cite{Tesanovic1,Tesanovic2} and by Sudb{\o} and collaborators \cite{sudbo99,sudbo4,sudbo5}. 
Although our model and that of Te{\v s}anovi{\' c} \cite{Tesanovic2} 
yield free energies that are formally very similar, the two models differ 
in one important respect. 
In Ref. \onlinecite{Tesanovic2}, the matter field $\Phi$ that describes the 
relativistic 
quantum bosons corresponds to a {\it fictitious} 
vortex system of {\it zero total vorticity}. The field-induced vortices
are separated out of the field $\Phi$. In our model, in contrast, 
the boson field $\Phi$  
corresponds to the {\it actual} flux-line 
system, including both field-induced and spontaneously generated vortices. 
The externally applied magnetic field ${\bf H}$ which yields a 
non-zero vorticity enters explicitely in the free energy and controls
the net mean flux threading the superconductor. It corresponds to a 
chemical potential for the bosons which yields a non-zero average charge. 
Furthermore, Te{\v s}anovi{\' c}'s fictitious vortex loops have 
a short-range scalar steric repulsion which enters as a $\Phi^4$ coupling in
the action. This is in addition to the usual screened 
magnetic interaction and  results from the singular gauge 
transformation which gives rise to the fictitious vortex loops. Such a 
steric repulsion does not exist among the {\it actual} vortex 
loops \cite{Blatter,BrandtRev} and therefore it is absent in our model.
In our action, vortices only interact through the long range magnetic 
interaction (screened by the Chern-Simons term) 
which is cutoff at distances $\leq \xi$ to ensure a finite
repulsive energy barrier at short distances.

An important property of the flux-line array which provides a direct measure 
of vortex correlations along the direction of the applied external field 
is the tilt modulus. A drawback of the hydrodynamics of directed 
flux-lines familiar from the literature\cite{paper1} is that it has been 
impossible to
separate the renormalization of 
the single-vortex part of the tilt modulus (related to the single-line 
tension) from that of the compressional 
part. The  hydrodynamic model that we present here allows us to 
probe the renormalization of the single-line tension through the 
autocorrelator of the $t_z^L$ field. This is an interesting direction 
for future work as the vanishing of the effective line tension is 
considered to be a signature of the $\Phi$-transition suggested by 
Te{\v s}anovi{\' c} \cite{Tesanovic2} and by Sudb{\o} and 
collaborators  \cite{sudbo99,sudbo4,sudbo5}. 

Finally, the hydrodynamic model of {\it directed} flux-line liquids has successfully been 
used to evaluate the renormalization of the tilt modulus 
from columnar defects parallel to 
the external magnetic field \cite{paper2}. This type of quenched disorder
suppresses the wandering of the flux-lines away from their 
average direction and it renders 
the directed line approximation appropriate. In contrast,
other types of quenched disorder,  namely point 
defects or splayed columnar defects with large splay, enhance vortex 
wandering, so that the resulting vortex liquid  can no longer be 
described as a liquid of directed lines.
The hydrodynamic model introduced here is particularly appropriate for 
studying the effect of 
this kind of disorder, which can be modelled by a random potential coupled 
to the field $\rho$. This is another interesting direction for 
future work.
\vspace{0.2in}

This work was supported by the National Science Foundation through Grant No. DMR97-30678. It is a great pleasure to acknowledge valuable discussions with Steve Teitel. PB acknowledges the hospitality of the Physics Department of the Technical University of Munich where part of this work was completed.

\begin{center}
{\bf APPENDIX: ANISOTROPIC SUPERCONDUCTORS}  
\end{center}

Our discussion has so far been limited to isotropic materials.
High-$T_c$ superconductors are,
however, layered materials where the anisotropy can play an important 
role and change substantially vortex behavior. For this reason in this section we generalize our model 
to include finite anisotropy. As usual, anisotropy is incorporated in the Ginzburg-Landau
free energy via an anisotropic effective mass tensor which leads to different 
values for the penetration and coherence lengths in the $ab$ plane and in the $c$ direction \cite{Blatter}.
 
We first consider a uniaxial superconductor in an external field
${\bf H}$ applied along the $c$ axis chosen as the $z$ direction.
The energy of a single vortex line with position parametrized  
as ${\bf r}=\big[{\bf r}_\perp(z),z\big]$ (${\bf r}_\perp(z)$ can be a multivalued function of $z$ to allow for the possibility of overhangs) wandering 
between points $a$ and $b$ is given by
\begin{equation} 
\label{singlelineanis} 
\tilde{U}_{self}\approx\epsilon_1\int_a^b \sqrt{(dz)^2+\frac{1}{p^2}(d{\bf r}_{\perp})^2}\;, 
\end{equation} 
where $p=\lambda_c/\lambda_{ab}$ is the familiar anisotropy parameter.
In copper-oxide high-$T_c$ materials, $p>>1$.
By comparing Eq. (\ref{singlelineanis})  to the action of a free relativistic particle given in 
Eq. (\ref{relact}), we see that a vortex line in a uniaxial material 
can be interpreted as the world line of a relativistic particle
which moves in an ``anisotropic space-time'' with rescaled spatial 
coordinates, ${\bf r}_{\perp}\rightarrow {\bf r}_{\perp}/p\;$. 
At the single-vortex level, anisotropy 
effectively {\it enhances} the 
``relativistic'' behavior, as it reduces the energy per unit length
associated with transverse vortex fluctuations, as expected for a layered material.
In this context, it may be tempting to interpret anisotropy 
as responsible for an effective reduction of the speed of light in the
relativistic $2D$  boson problem
with $c\rightarrow c/p$. This interpretation is, however, misleading as
it does not carry through when interactions are included.
In the interaction part
of the vortex-line free energy, anisotropy does not simply lead to a rescaling
of the transverse coordinates. This is apparent from the fact that the collective
part of the wavevector-dependent elastic constants of a vortex lattice in a uniaxial material
is not simply obtained from the corresponding elastic constants of an isotropic material with
the replacement $q_\perp\rightarrow q_\perp p\;$.

In the interaction and the free field part of the boson model, the role of anisotropy is simply
that of allowing for different scalar and transverse interactions among the bosons,
precisely as originally proposed by Feigel'man and collaborators \cite{feig}. Note that the aforementioned parts in our relativistic boson action are formally the same as the corresponding parts in Feigel'man's boson action - our model differs in the free particle part only. The relativistic boson action that maps onto the free energy
of interacting vortex lines and loops in a uniaxial superconductor
with an external field applied along the $c$ axis is given by
\end{multicols} 
\begin{eqnarray} 
\label{Relanis} 
\tilde{\cal S}_r[\Phi, \Phi^*, {\bf a}, {\bf A}]= 
  \int_0^{\beta \hbar} d\tau  \int d^2{\bf r}_{\perp}\bigg\{& &
\frac{mc^2}{2}\Phi\Phi^* +\frac{1}{2m_bc^2}\big[(-i\hbar\partial_\tau+e_ba_0+i\mu_r)\Phi\big] 
      \big[(i\hbar\partial_\tau+e_ba_0+i\mu_r)\Phi^*\big]\nonumber\\ 
& & +\frac{p^2}{2m_b}|(-i\hbar{\bf \nabla}_\perp+\frac{e_b}{c}{\bf
a}_\perp)\Phi|^2 + \tilde{\cal L}_F[{\bf a}, {\bf A}]\bigg\}\;, 
\end{eqnarray} 
where 
\begin{eqnarray} 
\label{Lfieldanis} 
\tilde{\cal L}_F[{\bf a}, {\bf A}]= \frac{1}{8\pi}\bigg\{p^2({\bf 
\nabla}_{\perp}\times{\bf a}_{\perp})^2+\big[\hat{\bf 
z}\times(\frac{1}{c}\partial_\tau{\bf a}_\perp-{\bf \nabla}a_0)\big]^2 +\frac{2i}{{\lambda}_b}({\bf 
\nabla}\times{\bf a})\cdot{\bf A} 
+({\bf                                  
\nabla}\times{\bf A})^2\bigg\}\;.  
\end{eqnarray} 
\begin{multicols}{2} 
\noindent We stress that in this mapping the boson mass, $m_b$,
corresponds to the vortex {\it line energy}, $\epsilon_1$, precisely as indicated in Table 1.
This is indeed the appropriate interpretation as it is made apparent by comparing the 
single-vortex energy given in Eq. (\ref{singlelineanis}) to the  action
of a relativistic boson, given in Eq. (\ref{relact}).
In contrast, in all the previous literature, and particularly in the work by
Nelson and coworkers, it is  the tilt
energy per unit length, $\tilde{\epsilon}_1=\epsilon_1/p^2$, that is interpreted as the boson mass.

Finally, by using the methods described in Section III and the translation Table 1,
one can immediately obtain the hydrodynamic free energy 
of a liquid of arbitrarly fluctuating vortex lines and loops
in a uniaxial superconductor with ${\bf H}\parallel c$.
It is given by
\end{multicols} 
\begin{eqnarray} 
\label{vtRelanis} 
\tilde{\cal F}_r[{\bf t}, \rho]=& &
  \int_{\bf r}\bigg\{
     \frac{(k_BT)^2}{8\tilde{\epsilon}_1 \rho}({\bf \nabla}_{\perp}\rho)^2
    +\frac{(k_BT)^2}{8\epsilon_1 \rho}(\partial_z\rho)^2
    +\frac{\tilde{\epsilon}_1}{2 \rho}({\bf t}_{\perp})^2 
    +\frac{\epsilon_1}{2 \rho}(t_z)^2 
    +\frac{\epsilon_1}{2}\rho - {\mu_r}t_z
  \bigg\} \nonumber\\  
& & +\frac{1}{2\Omega}\sum_{\bf q}\bigg\{
    \frac{4\pi\epsilon_0\tilde{\lambda}_\perp^2}
           {1+q_z^2{\tilde\lambda_\perp}^2+q_\perp^2p^2{\tilde\lambda_\perp}^2}
           |{\bf t}_{\perp}({\bf q})|^2   
    +\;\frac{4\pi\epsilon_0\tilde{\lambda}_\perp^2(1+q^2p^2{\tilde\lambda_\perp}^2)}
            {(1+q^2{\tilde\lambda_\perp}^2)(1+q_z^2{\tilde\lambda_\perp}^2
            +q_{\perp}^2p^2{\tilde\lambda_\perp}^2)}
     |t_z({\bf q})|^2  
  \bigg\}\; 
\end{eqnarray} 
\begin{multicols}{2} 
\noindent with $\tilde{\epsilon}_1=\epsilon_1/p^2$ and the familiar constraint ${\bf \nabla}\cdot {\bf t}=0\;$. 

\end{multicols}

\bibliography{rel}

\begin{thebibliography}{10}
\bibitem{charge_note}
Here all bosons have the same charge, either positive or negative. In contrast,
  the relativistic boson system discussed below consists of a mixture of
  positively and negatively charged bosons.
\bibitem{foot_mu}
The chemical potential, $\mu$, customarily defined in non-relativistic
  many-body theory, is related to our $\mu_r$ by $\mu=\mu_r-mc^2$ (for
  $\mu_r\geq 0$, chosen here to correspond to particles). The equilibrium
  solution is then given by $\rho_0=0$ for $\mu\leq 0$, and
  $\rho_0=\mu/(4\pi\lambda_b^2e_b^2)$ for $\mu> 0$.
\end{thebibliography}

\end{document}